\documentclass[%
 aip,
 sd,%
 amsmath,amssymb,
 reprint,%
]{revtex4-1}

\usepackage{graphicx}
\usepackage{dcolumn}
\usepackage{bm}
\usepackage{color}
\usepackage{braket}

\begin{document}

\preprint{AIP/123-QED}

\title[]{Muon probes of temperature-dependent charge carrier kinetics in semiconductors}

\author{K.~Yokoyama}
\email{koji.yokoyama@stfc.ac.uk}
\affiliation{
ISIS, STFC Rutherford Appleton Laboratory, Didcot, OX11 0QX, United Kingdom
}

\author{J.~S.~Lord}
\affiliation{
ISIS, STFC Rutherford Appleton Laboratory, Didcot, OX11 0QX, United Kingdom
}
\author{P.~W.~Mengyan}
\affiliation{
Department of Physics, Northern Michigan University, Marquette, Michigan 49855, USA
}
\affiliation{
Department of Physics and Astronomy, Texas Tech University, Lubbock, Texas 79409-1051, USA
}

\author{M.~R.~Goeks}
\affiliation{
Department of Physics, Northern Michigan University, Marquette, Michigan 49855, USA
}

\author{R.~L.~Lichti}
\affiliation{
Department of Physics and Astronomy, Texas Tech University, Lubbock, Texas 79409-1051, USA
}

\date{\today}

\begin{abstract}
We have applied the photoexcited muon spin spectroscopy technique (photo-$\mu$SR) to intrinsic germanium with the goal of developing a new method for characterizing excess carrier kinetics in a wide range of semiconductors. 
Muon spin relaxation rates can be a unique measure of excess carrier density and utilized to investigate carrier dynamics. The obtained carrier lifetime spectrum can be modeled with a simple diffusion equation to determine bulk recombination lifetime and carrier mobility. 
Temperature dependent studies of these parameters can reveal the recombination and diffusion mechanism.
\end{abstract}

\maketitle

The kinetics of excess charge carriers in semiconductors are crucial in determining the performance of electronic devices. 
For high efficiency photovoltaic cells, the majority of excess carriers need to diffuse across the $p$-$n$ junction before they recombine. 
Carrier mobility directly affects device switching speed, a crucial factor not only for computer chips but also for power electronic devices. 
Photoinjected carriers drive chemical reactions in photocatalytic agents, such as  TiO$_{2}$ nano-particles. 
A precise understanding of carrier transport properties is quite valuable for material physics research and in device optimization. 
Researchers are therefore committed to developing measurement techniques such as photoluminescence and transient absorption/reflection spectroscopy.~\cite{Schroder} 
In silicon industries, the microwave-detected photoconductivity method is widely used to measure excess carrier lifetime and characterize wafers. 
On the other hand, it is often essential to combine methods in order to obtain a comprehensive picture as done in recent years with the appearance of a number of novel functional materials such as perovskite-structured compounds and wide-bandgap semiconductors for solar cell and power device applications, respectively.

With the goal of developing a new method applicable to a wide range of semiconductors, we previously used intrinsic Si as a proof-of-concept system, applied the photoexcited muon spin spectroscopy technique (photo-$\mu$SR) and successfully measured the excess carrier lifetime and diffusion constant for intrinsic Si.~\cite{YokoyamaPRL} 
When a positively charged (anti)muon~$\mu^+$ (elementary particle, charge of~+$e$, spin of~1/2 and~1/9 the mass of a proton) with an initial energy of 4~MeV is implanted in a material, the muon thermalizes over several hundred micrometers making it an ideal {\it bulk} probe of material.~\cite{Blundell} 
In many semiconductors the implanted muon captures an electron and forms a hydrogen-like atom, muonium $(\text{Mu}=\mu^{+}+\text{e}^{-})$, which can undergo spin and carrier exchange interactions with free electrons and holes.~\cite{Chow} 
If initially in a triplet state, $\ket{\mu \Uparrow \text{e} \uparrow}$, Mu will be depolarized upon conversion into the singlet state, $\ket{\Uparrow \downarrow}$, due to hyperfine (HF) interaction of $\mu^+$ and e$^-$. 
Higher carrier concentrations cause more carrier cycling thereby converting more $\ket{\Uparrow \uparrow}$ states into $\ket{\Uparrow \downarrow}$ states resulting in faster spin relaxation. 
Mu is a defect center and can induce carrier recombination by itself, but since the number of implanted muons is on the order of $10^{3}~\mu^{+}$/pulse, its density and contribution to carrier kinetics is negligible. 
By combining optical excitation and $\mu$SR methods, one can then pump carriers and probe their dynamics using muons in a contact-free environment.~\cite{YokoyamaPRL} 
Thanks to the high penetration of muons into matter, a sample can be contained in a cell for cooling down to cryogenic temperatures or heating up to an annealing condition.  
It is thus straightforward to perform temperature and injection level dependent measurements, which often give us clues for impurity states~\cite{ReinBook} and can significantly contribute to the characterization of new materials.

In this Letter, we extend the photo-$\mu$SR pump-probe technique\cite{YokoyamaPRL, YokoyamaRSI} to germanium, another representative semiconductor, and focus on a temperature study to demonstrate how much information on excess carriers is available by utilizing this localized probe and new method. 
$\mu$SR work on Ge has been established now for a few decades (see for example refs~~\cite{Patterson, Cox, Lichti1999}). 
Illumination of Ge is known to affect the time-evolution of the muon's spin-polarization in a way that is similar to what is observed in Si.~\cite{KadonoEtAl} 
From an application point of view, Ge was supplanted by Si in the early stage of the semiconductor history because of the (i)~lower abundance, (ii)~higher cost, and (iii)~lower bandgap energy resulting in a lower maximum operating temperature. 
However, there are attempts in some device applications to replace Si with Ge by virtue of its faster carrier transport.~\cite{Ye}

\begin{figure}
\includegraphics{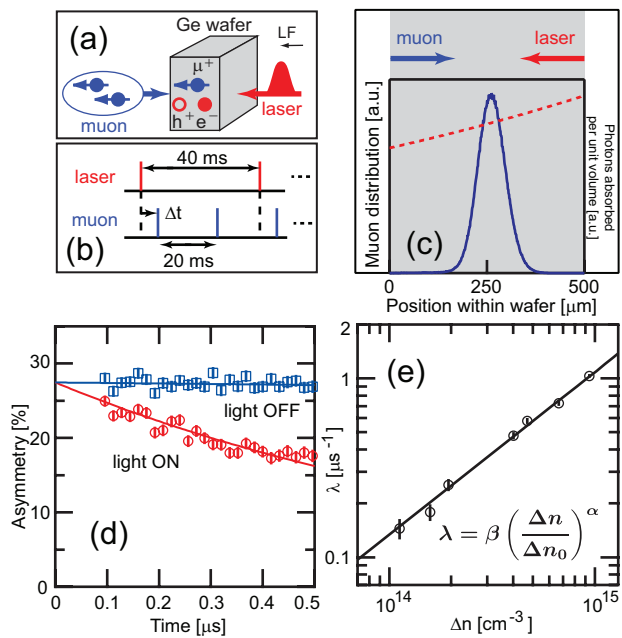}
\caption{\label{fig:Exp}
(Color online) 
(a)~Schematic diagram of the experimental geometry. 
(b)~Timing diagram of laser and muon pulse. Pulse duration (FWHM) of the laser and muon pulse are 10 and 70~ns, respectively. 
(c)~Calculated muon distribution (solid blue line) and photon flux (red broken line) as a function of position within the sample measured from the surface on which the $\mu^{+}$ are incident. 
(d)~Representative $\mu$SR time spectra for light OFF (blue open squares) and ON (red open circles) and their fit (solid lines). $2.5\times 10^{6}$~muon decay events are averaged for each spectrum. Fit parameters (see main text) are $A(0) = 27.45 \pm 0.02~\%$, $\lambda '= 0.0250 \pm 0.0002~\mu \text{s}^{-1}$, and $\lambda = 1.02 \pm 0.02~\mu \text{s}^{-1}$. 
(e)~Calibration curve for $T = 295$~K, plotting $\lambda$~vs~$\Delta n$. The fit parameters are $\alpha = 0.91 \pm 0.03$, $\beta = 1.03 \pm 0.02~\mu \text{s}^{-1}$ and $\Delta n_0 = 9.4\times 10^{14}~\text{cm}^{-3}$.}
\end{figure}

The photo-$\mu$SR experiment on Ge presented here\cite{DOI} was carried out using the HiFi spectrometer at the ISIS Neutron and Muon Source at the STFC Rutherford Appleton Laboratory in the UK. 
Complete details of the experimental apparatus are found in refs~~\cite{YokoyamaRSI, Lord}. 
Briefly, as shown in Fig.~\ref{fig:Exp}(a), 100\% spin-polarized muons are incident on one side of the sample while pump light illuminates the other side. 
Here, the sample is a 2-inch diameter, 500-$\mu$m thick, intrinsic, single crystal Ge wafer ($n$-type,  $R>50~\Omega\cdot$cm, both sides polished) with the $\langle 111 \rangle$ axis perpendicular to the surface. 
A magnetic field is applied parallel to the initial $\mu^{+}$~spin direction (longitudinal field, LF), which is opposite the $\mu^{+}$~momentum direction. 
Aluminium foils are used (as a degrader) to position the average muon implantation depth in the center of the sample. 
This can be accurately predicted with the help of \textit{musrSim}~\cite{Sedlak} (a Monte Carlo simulation package based on GEANT~4) using the known incoming $\mu^{+}$ momentum along with the amount and density of materials in the beam. 
Fig.~\ref{fig:Exp}(c) shows the optimized muon stopping profile with a FWHM $\approx$100~$\mu$m. 
Once implanted and fully thermalized, Mu does move within the Ge crystal lattice,~\cite{Patterson, Lichti1999} but the diffusion constant is, at most, 10$^{-3}$~cm$^{2}$/s and therefore negligibly small when considering the timescale and size of the system. 
Also drawn in Fig.~\ref{fig:Exp}(c) is the exponential decay of photon flux due to absorption noting that the photons arrive at the surface (500~$\mu$m) opposite to that of the muons (0~$\mu$m). 
The laser light is generated by an optical parametric oscillator~\cite{YokoyamaRSI} whose output wavelength is selected depending on the sample temperature $T$ such that the absorption coefficient is within 3$\sim$10~$\text{cm}^{-1}$.~\cite{Macfarlane} 
For instance at room temperature, wavelengths of 1825 and 1760 nm give the absorption coefficients 3 and 10~$\text{cm}^{-1}$, respectively. 
Injected excess carrier density $\Delta n$ is calculated for the muon position using the measured pulse energy of laser light and its spot size ($\approx$7~cm$^2$). 
As shown in Fig.~\ref{fig:Exp}(b), the muon and laser pulses operate at 50 and 25~Hz, respectively, with a tunable delay of $\Delta t$ between them, so that light~ON and~OFF spectra are measured alternately. 
Fig.~\ref{fig:Exp}(d) shows example light ON and OFF muon spectra measured at 295~K with 1785~nm laser light, which pumps $\Delta n =9.4 \times 10^{14}$~cm$^{-3}$ at $\Delta t = 0$. 
The light~ON spectrum is fitted to $A(t) = A(0)e^{-(\lambda ' + \lambda) t}$, where $A(t)$ is the measured time-domain muon asymmetry, $\lambda '$ is the relaxation rate for light~OFF, and $\lambda$ is the rate induced by the carrier injection. 
During this short window $\Delta n$ is assumed to be constant, hence the obtained rate $\lambda$ uniquely tags this~$\Delta n$.~\cite{Note1} 
The pump power is then attenuated with neutral density filters to make a calibration curve as shown in Fig.~\ref{fig:Exp}(e), which allows $\Delta n$ to be calculated using the fit function shown in Fig.~\ref{fig:Exp}(e) for a measured $\lambda$.~~\cite{YokoyamaPRL}

Shown in Fig.~\ref{fig:Exp}(d) are representative muon asymmetry spectra from measurements where a 1.0~Tesla LF is applied to the sample to partially decouple the Mu HF interaction and hence adjust the relaxation rates $(\lambda)$ to be within an appropriate regime for the calibration curve.~\cite{YokoyamaPRL} 
An appropriate LF should be selected for a given temperature because $\lambda$ can be different even with the same $\Delta n$. 
Similar to Si, the positive and neutral muonium centers are supported at the bond-center site (Mu$_{BC}^{+}$ and Mu$_{BC}^{0}$) and the negative and neutral centers are supported at the interstitial tetrahedrally coordinated site (Mu$_{T}^{-}$ and Mu$_{T}^{0}$).~\cite{Patterson, Cox} 
Upon implantation, muons are distributed between these states with a ratio that varies with temperature and the Fermi energy. 
Carrier injection affects the dynamics of these centers through interactions involving spin, carrier and site exchanges that form a complex network. 
The depolarization of the muon spin is affected by cycles involving excess carriers and the Mu HF interaction (mainly in Mu$_{T}^{0}$). 
The details relating to the mechanisms involved in this dynamic network are beyond the scope of this Letter, but part of our ongoing larger-scale study. 
Here we empirically utilize the fact that~$\lambda$ provides a good measure of~$\Delta n$.

\begin{figure}
\includegraphics{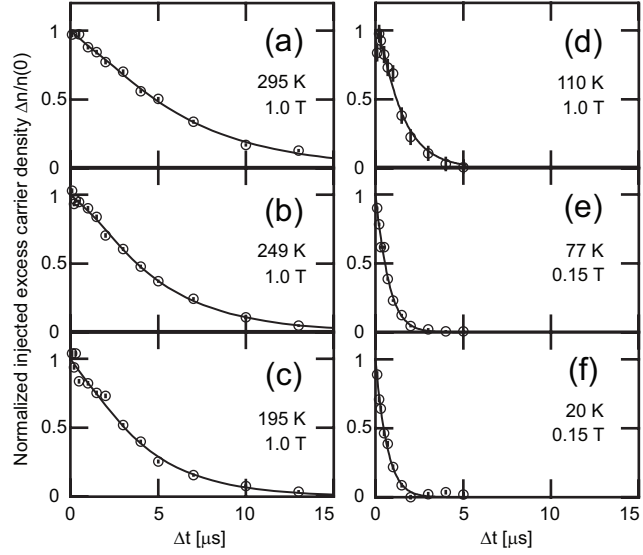}
\caption{\label{fig:LTS} 
Excess carrier lifetime spectra for (a)~295~K, (b)~249~K, (c)~195~K, (d)~110~K, (e)~77~K, and (f)~20~K. 
Applied fields (LF) are shown within each figure. 
Note that the calibration curve needs to be measured at each temperature even if the same field is used (different fractions of Mu states and light OFF transition rates). 
Solid lines represent a fit to the model (see text). 
For simplicity, the y-axis in each is the injected excess carrier density $\Delta n$ normalized to $\Delta n(0)$. 
The same injection level, $\Delta n(0) \approx 10^{15}~\text{cm}^{-3}$, was used in all measurements.}
\end{figure}

After setting~$\Delta n$ to the maximum value in Fig.~\ref{fig:Exp}(e), $\Delta t$ sweeps through the lifetime period to measure~$\lambda (\Delta t)$, which is then converted into $\Delta n (\Delta t)$ using the calibration curve. 
Fig.~\ref{fig:LTS} shows the carrier lifetime spectra for six temperatures from 20~K to 295~K. 
The spectra can be modeled with a 1-dimensional diffusion equation for $\Delta n(z,t)$,

\begin{eqnarray}
D\frac{\partial^{2}\Delta n}{\partial z^{2}}-\frac{\Delta n}{\tau_{b}}=\frac{\partial \Delta n}{\partial t}
\;
\label{eq:bulk},
\end{eqnarray}

\noindent
where $D$ is the effective carrier diffusion constant, $\tau_{b}$~is the bulk recombination lifetime, and $z$~is the position within the sample along the axis of the muon and laser beams. 
The position of the surface on which the muons are incident is set as~$z=0$ (laser incident on the opposite surface; Fig.~\ref{fig:Exp}(c)). 
The wafer has been lapped and mechanically polished without any follow-up passivation processes and so the surface velocity is expected to be very high ($>10^5$~cm/s). 
In Si, surface velocities faster than $10^4$~cm/s do not exhibit any dependencies on temperature, injection level, or resistivity.~\cite{Willander} 
We therefore impose boundary conditions for the surfaces of a wafer with thickness $d$ to be $\Delta n(0,t) = \Delta n(d,t) = 0$, and analytically solve Eq.~(\ref{eq:bulk}) with the initial condition~$\Delta n(z,0) = \Delta n(0)$. With $D$, $\tau_{b}$, and $\Delta n(0)$ as fit parameters, the solid lines in Fig.~\ref{fig:LTS} show a fit to $\Delta n(d/2,t)$ for each lifetime spectrum. 
The fitting curves reproduce all of these data within error bars, indicating these data are consistent with the diffusion model. 

\begin{figure}
\includegraphics{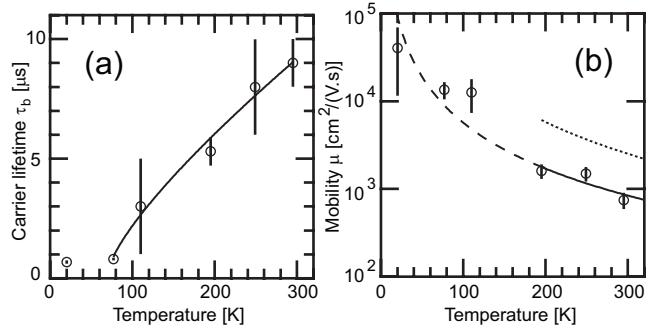}
\caption{\label{fig:Diff} 
(a) Temperature dependence of bulk carrier lifetime obtained from Fig.~\ref{fig:LTS}. 
The solid line is a fit to $\tau_b(T) = \tau_{0} + c^{\prime}(T - T^{\prime})^{p}$ for $T > T^{\prime}$ with fixed $\tau_{0} = 0.7~\mu\text{s}$, assuming that the observed $\tau_b$(20~K) =  0.68 $\pm$ 0.07~$\mu$s provides $\tau_0$. 
Obtained fit parameters are: $T^{\prime}$ = 75 $\pm$ 7 K,  $c^{\prime}$ = 0.12 $\pm$ 0.10, and $p$ = 0.8 $\pm$ 0.2. 
(b) Temperature dependence of carrier mobility. 
Solid line denotes a fit to $\mu(T) = c^{\prime \prime} T^q$ [cm$^2$/(V$\cdot$s)] for $T > $ 195 K. 
Fit parameters are: $c^{\prime \prime}$ = (1.7 $\pm$ 3.0) $\times$10$^7$ and $q$ = $-$1.7 $\pm$ 1.0. 
The curve is extrapolated down to 20~K (broken line).
Dotted line represents $\mu_{\text{a}}(T)$ calculated from the literature values, $\mu_\text{e}(T) = 4.9 \times 10^7 T^{-1.66}$ and $\mu_\text{h}(T) = 1.1 \times 10^9 T^{-2.33}$ (see text).~\cite{Morin}}
\end{figure}

Based on the analyses demonstrated in Fig.~\ref{fig:LTS}, we now discuss temperature dependence of $\tau_b$ and carrier mobility $\mu$, which is calculated from an electrical mobility equation, $D = \mu k_B T/e$, where $k_B$ is the Boltzmann constant and $e$ the electrical charge of an electron. 
Fig.~\ref{fig:Diff}(a) shows that $\tau_{b}$ decreases monotonically with decreasing temperature to 77~K and then seems to stay constant through at least 20~K. 
Comparing with various capture mechanisms in impurity sites,\cite{ReinBook} this feature is consistent with a study on high-resistivity Si with Fe as an intentionally doped, deep-level recombination center.~\cite{Hangleiter_ex} 
Hangleiter has explained this using a model called excitonic Auger recombination, which attributes the fast carrier decay to efficient recombination by exciton formation.~\cite{Hangleiter_th} 
When a carrier in a free exciton is captured and recombines in an impurity site, it always has its pair particle in the vicinity, which can take excess energy. 
Therefore, the excitonic Auger capture mechanism takes place very efficiently in the defect centers. 
Since the associated capture cross-section~$\sigma$ depends on the thermal ionization of excitons, $\sigma$ becomes temperature independent below a threshold~$T^{\prime}$, where all carriers form free excitons. 
The model predicts $\sigma(T)$ to be constant for $T< T^{\prime}$ and to follow a power law, $\sigma(T) \propto T^{-p}$, for $p>0$ and $T> T^{\prime}$. 
Because $\tau_{b} \propto \sigma^{-1}$, we expect $\tau_{b}(T)$ to also follow a power law $T^{p}$. 
As shown in Fig.~\ref{fig:Diff}(a), the model gives an excellent fit with $p$ = 0.8, implying that the capture efficiency by deep centers decreases with $T^{-0.8}$ along with exciton ionization in the present system (Ge).
The obtained threshold temperature $T^{\prime}$ = 75 $\pm$ 7~K is comparable with 60~K measured in Si.~\cite{Hangleiter_ex} 
This is consistent with the fact that the free exciton binding energy for Si and Ge are also comparable.~\cite{Note2}

Fig.~\ref{fig:Diff}(b) shows that carrier mobility monotonically increases with decreasing temperature, which is characteristic of lattice scattering. 
Previous studies report that the conribution by lattice scattering can be described with a power law temperature dependence for both electrons and holes before impurity scattering becomes more significant below 100~K.~\cite{Morin} 
In the present system the ambipolar mobility characterized by $\mu_{\text{a}} = 2\mu_{\text{e}}\mu_{\text{h}}/(\mu_{\text{e}}+\mu_{\text{h}})$ should describe the behavior in the high temperature range, where electrons ($\mu_{\text{e}}$) and holes ($\mu_{\text{h}}$) diffuse together due to the Coulomb interaction.~\cite{Neamen} 
Therefore, the data between 195 and 295~K are fitted with a power law and yield $\mu_{\text{a}}(T) \propto T^{1.7 \pm 1.0}$. 
As shown in Fig.~\ref{fig:Diff}(b), despite the relatively large error, the obtained power law is consistent with the calculation, $\mu_{\text{a}}(T) \propto T^{2.05}$. 
The lower value of mobility compared with the calculated curve suggests that our sample contains more impurities. 
As expected, the ambipolar model is no longer valid in the low tempearture range, where exciton diffusion and impurity scattering should be dominant. 
The measurement gives the free exciton diffusion constant D$_{\text{ex}}$ = 70 $\pm$ 50~cm$^2$/s in 20~K, which is lower than the previous study reporting D$_{\text{ex}}$ = 300~cm$^2$/s, probably due to the higher impurity concentration. \cite{Culbertson} 

In summary, we have successfully applied the photo-$\mu$SR technique to intrinsic Ge and measured the temperature dependences of carrier lifetime and mobility with a simple diffusion model. 
The lifetime measurement has identified the main recombination mechanism as the excitonic Auger process in deep centers. 
The temperature dependence of carrier mobility is found to follow ambipolar diffusion in the high temperature range, but is dominated by excitons in the low temperatures. 
Results from this new photo-$\mu$SR method are consistent with the previous results and prove that the photo-$\mu$SR method can correctly capture carrier kinetics. 
The photo-$\mu$SR technique is unique in that the muon is a spatially well-defined probe that enables us to investigate the entire carrier dynamics instead of measuring a single parameter. 
Although in the present study the muon distribution is centered in the wafer, it can be shifted to one side of the wafer by adding more degraders in the muon beam. 
Measurements using different depths within a wafer gives a steric description of carrier kinetics and may enable the study of surface conditions (e.g.~surface velocity, space charge distribution) on a passivated surface. 
Last but not least, Mu can be implanted in many semiconductors and insulators where Mu will interact with excess carriers in a way that is similar to what we have seen in Si and Ge. 
Therefore, this technique is (in principle) applicable to a wide range of semiconductor systems, so long as the timescale of carrier recombination is on the order of $\mu$s or longer. 
This technique may not work very well when the timescale is on the order of ns, such as what is typically found in direct-gap semiconductors. 
However, as long as there is a fast carrier-induced depolarization,~\cite{YokoyamaPP} a carrier measurement may be possible by considering a convolution of the carrier-induced depolarization and temporal profile of the laser pulse.

This work was supported by the Science and Technology Facilities Council in the UK. 
Additionally, support is acknowledged from the Texas Research Incentive Program (PWM, RLL) and the NMU Freshman Fellows Program (MRG). 
We wish to acknowledge the assistance of a number of technical and support staff in the ISIS facility.


\end{document}